# Statistical Assessment of Safety Levels of Railway Operators

**Jens Braband[a], Hendrik Schäbe[b]**

[a]Siemens Mobility GmbH, Braunschweig, Germany

[b]TÜV Rheinland InterTraffic GmbH, Cologne, Germany

**Abstract:** Recently the European Union Agency for Railways (ERA) has received a mandate for "the development of common safety methods for assessing the safety level and the safety performance of railway operators at national and Union level" [2]. Currently, several methods are under development. It is of interest how a possible candidate would behave and what would be the advantages and disadvantages of a particular method. In this paper, we study a version of the procedure. On the one hand side we analyze it based on the theory of mathematical statistics. As a result, we present a statistically efficient method the rate-ratio test based on a quantity that has smaller variance than the quantity handled by the ERA. Then, we support the theoretical results with the help of a simple simulation study in order to estimate failure probabilities of the first and second kinds. We construct such alternative distributions which the decision procedure cannot distinguish. We will show that the use of procedures that are optimal in the sense of mathematical statistics combined with the use of a characteristics that has small spread – here the number of accidents – is advantageous.

## 1. Introduction

In the Commission Decision [1], a set of so-called NRV (national reference values) for each member state of the EU is estimated from accident data for several years. The accident data is measured in FWSIs (fatalities and weighted serious injuries) for different categories of persons: passengers, employees, level-crossing users, others, unauthorized persons, society as a whole. Recently the European Union Agency for Railways (ERA) has received a mandate for "the development of common safety methods for assessing the safety level and the safety performance of railway operators at national and Union level" [2].

The work is still ongoing [3] but the basic approach is established and as the result will have massive impact on railway operators in Europe it is reasonable to analyze the approach at a very early level. In this paper we focus on the assessment of the so-called safety level of operators only, which is defines as the "level of occurrences of eligible events … estimated and assessed by the method defined …". The approach for this assessment is a statistical approach while safety performance is defined as a maturity level for the operator's risk management.

The question to be answered each year for each operator and different event categories can be defined by the following hypothesis: "$H_0$: The safety performance of the operator has not deteriorated with respect to a particular reference period.". Note that also comparisons between different operators shall be made. This type of hypothesis, i.e. a certain parameter has not deteriorated, is well-known for problems of statistical quality control. There, parameters such as e.g. lengths, widths etc. are monitored. A sampling technique is applied to judge whether the process produces items that are out of specification. In quality control often a normal distribution with mean value $\mu$ and a spread $\sigma$ is supposed to exist for the parameter to be monitored. Then, limits are computed and if a certain number of items are outside the limits, the process is judged to be out of control. Details can be found in many books on statistical quality control, as e.g. [4]. In order to monitor the process, quality control charts are used, where the measurement values are registered together with a running number or the time the measurement was carried out. Then, on the control charts the warning limits and the action limits are drawn according to the rules applied. Later, these control charts have been refined and computerized.

So, the problem treated by the ERA in its mandate [1] is like that of the well-known control charts. However, in contrast to the very specific algorithm prescribed by the ERA for national safety assessment [1] in the new approach ERA seems not to define any particular method. The discussion of the general approach and proposed methods is subject of the present paper.

In section two we will describe the approach used by the ERA to monitor their indicators. In the third section we will provide a theoretical model for the counting processes that are laying behind the data considered by the ERA. Section 4 presents a survey on previous work. The fifth section is dedicated to severity distributions. In section six we propose the rate-ration test to evaluate accident frequencies and to discuss different approaches. The last section is dedicated to conclusions.

**2.     The Approach Used by the ERA**

"The general objective [of the ERA proposal] is to assess, with various tests, the extent to which an operator fulfils the requirement of maintaining and continuously improving railway safety." [3] Thus ERA wants to test the $H_0$ as defined above. This shall be done with respect to last year's safety level but also with respect to a 5 years reference period and against other comparable operator's safety level in the same year.

It is stated that "Statistical inference and tests shall be used to provide harmonized assessments to each operator." The outcome shall be classified either as "No deterioration", "Potential deterioration" or "Probable deterioration". The comparison shall be performed for different event frequencies which are normalized by the operation volume in the period under consideration.

For comparison of nationwide safety levels [1] ERA use a quite detailed process for which several statistical weaknesses are known [5], e. g. while the error probability of the first kind is very low, the error probability of the second kind is almost 1 in several scenarios. This means that while false alarms are minimized, trends that indicate deterioration of safety level are recognized very late. While this detailed procedure gives legal certainty for the nation states, it does not treat the uncertainty in the statistical data adequately.

For the new approach ERA seems to want to use more advanced statistical procedures but hesitates to fix them in the legal text [3]. This paper identifies and discusses some candidate methods.

## 3. General Model

There is broad agreement that a compound Poisson process with $N_t$ describing the number of accidents until time t and $S_1, S_2,\ldots$, an independent, identically distributed (i.i.d.) sequence of random variables describing the severity of the accidents, is an adequate model for the accident process. The main justification is that the Poisson process arises naturally from several limit theorems [6,13] and that accidents are rare events in railways.

The accidents occur with a rate of occurrence of $\lambda(t)$ at times $T_1, T_2, T_i$, etc., where i is the index of the accident. For each accident with index i there exists a certain "jump height" $S_i$, which is the severity. The severity can be measured as the number of fatalities (or equivalent fatalities) but may also be monetary. So, for a fixed interval [0,t], the process $X_t$ describes the cumulated number of fatalities and $N_t$ describes the accumulated number of accidents.

This compound Poisson process can be used to derive characteristics for collective risk.

So, under the null hypothesis, the observables $X_t$ are random variables that are distributed as

$$\sum_{i=1}^{N_t} S_i . \qquad (4)$$

With a constant accident rate $\lambda(t) = \lambda$

$$E(X_t) = \lambda t E(S_1) \text{ and } V(X_t) = \lambda t E(S_1^2). \qquad (5)$$

holds, where it has been assumed that the moments of $S_1$ exist. Given a sufficient number of accidents per year, limit theorems for the distribution of $X_t$ may apply, e.g. the central limit theorem.

**4. Previous Work**

Evans [7] has analyzed trends in railway safety in the EU and the UK and uses a similar Poisson model, which has been further analyzed by the authors [8].

Braband and Schäbe [9] have also analyzed for the general model classical parametric and non-parametric test and have proposed a Kolmogorov-Smirnov type test for the comparison of the safety level for two consecutive years. The test is non-parametric, but the evaluation of the test statistic relies on limit theorems, so that the test is not applicable for small sample sizes like they would often occur when assessing the safety level of single operators.

Andrasik [10] has assessed the safety level of operators for the Swiss national railway safety authority BAV by a Bayesian approach. He uses a particular a priori distribution that has been accepted by the BAV experts for corner cases where there are few events in the 4 years reference period and the 1-year target period.

| Number of events in reference period | Number of elements in reference period per year | Number of elements in target period (1 year) | | | | | | | |
|---|---|---|---|---|---|---|---|---|---|
| | | 0 | 1 | 2 | 3 | 4 | 5 | 6 | 7 |
| 0 | 0 | .50 | .74 | * .88 | + .95 | + .98 | + .99 | + 1 | + 1 |
| 1 | 0.25 | .44 | .68 | * .84 | +.93 | + .97 | + .99 | + 1 | + 1 |
| 2 | 0.5 | .38 | .62 | *.79 | +.90 | + .96 | + .98 | + .99 | + 1 |
| 3 | 0.75 | .32 | .56 | *.75 | *.87 | + .94 | + .97 | + .99 | + 1 |
| 4 | 1 | .27 | .50 | .70 | *.84 | + .92 | + .96 | + .98 | + .99 |
| 5 | 1.25 | .23 | .45 | .65 | *.80 | * .89 | + .95 | + .98 | + .99 |

Table 1  (1-p)-levels of the Andrasik test procedure

BAV assumes that if the probability is larger than 0.9 then a deterioration is probable (alert, marked by a + in table 1), and if the probability is larger than 0.75 then there is a potential deterioration (warning, marked by a * in table 1).

Andrasik includes severities into the evaluation and chooses a Binomial distribution

$$P(\text{Data}|p_f, p_s) = \binom{n_{f1} + n_{f2}}{n_{f1}} p_f^{n_{f1}} (1-p_f)^{n_{f2}} \binom{n_{s1} + n_{s2}}{n_{s1}} p_s^{n_{s1}} (1-p_s)^{n_{s2}}, \quad (1)$$

Here $n_{f1}$, $n_{f2}$, $n_{s1}$ and $n_{s2}$ denote the number of fatal and severe injuries in the periods 1 and 2, respectively. A Beta distribution is chosen as a prior.

Note that the Beta distribution is the so-called conjugate prior in this model [14]. At the one hand, it is the likelihood function of a fictitious previous sample and on the other hand, it can easily be combined with the likelihood function of the data, which makes algebra easy.

Here we want to note the following facts:

According to (1) the number of fatalities and severe injuries is assumed to be independent.

1. The number of fatal accidents and the number of serious accidents are described as independent statistical values, which does not reflect reality. Both values arise from the same events as collisions, derailments etc. and if the number of these events is reduced, the number of fatalities and injurie sis reduced in the same manner.

2. To base a statistical decision on a p-value of 0.75 might be questionable since with probability 25%, even if there is no deterioration of safety performance, a warning would be issued.

5.  **Discussion of Severity Distributions**

It is well-known from actuarial risk management, that loss distributions are positively skewed, usually have heavy tails and are often a discrete mixture of several distributions, e. g. Pareto, Weibull, etc...., see figure 2 for a typical example of the yearly data of a large railway operator. This makes the statistical evaluation quite sophisticated or complicated.

However, from a safety point of view each event of a particular class has a similar potential e. g. a derailment of a passenger train should be prevented, and only if this is not possible, then measures of passive safety e. g. enhancing crashworthiness would be implemented.

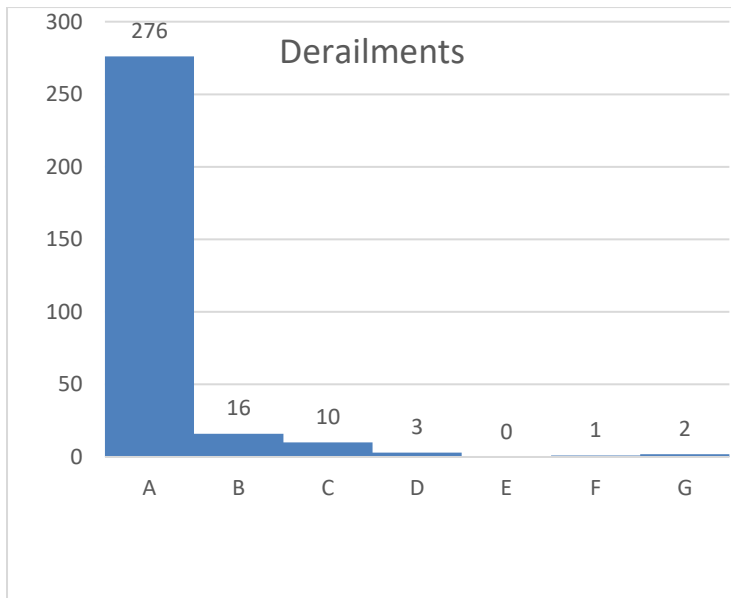

Figure 2 Number of derailments in loss classes A to G (G being the larges loss) for a large operator

But also, the large variance of the severity distributions may bias the results from comparison of safety levels and my lead to wrong conclusions. For example Evans [7] reports that since 1980, several years showed 3 severe level crossing accidents in Europe, the fatalities reported for these years were 33, 25, 15, 19, 27, 18, 9, 3, 11 and 13. In particular in 2016 3 fatalities were reported, while there were 11 in 2017 – both numbers of fatalities for the same number of accidents? Has level crossing safety deteriorated from 2016 to 2017? Or, if this occurred for a single operator, should we conclude that his safety level has deteriorated just because of the different (random) occupation of the cars that were involved?

A statistical analysis has been carried out on the ERA data [15]. Based on a set of 3152 accidents, it showed the following correlation coefficients between fatalities and sever injuries.

| Total fatalities | Passenger fatalities | LC User fatalities | Other fatalities |
|---|---|---|---|
| 0,63400074 | 0,63349783 | 0,96474093 | 0,95331197 |

Table 2  Correlations between fatalities and severe injuries

This computation supports the statement above.

Overall there are several good reasons not to include the severities in the evaluation of an operator's safety level and we will concentrate in the next chapter on the evaluation of accident frequencies only.

This statement can be supported by a small algebraic exercise. Let N denote the number of accidents and S be the number of fatalities in an accident for a certain time period. The N*S is the number of fatalities that occurred in that time period.

The variance of S*N is now:

$$Var(S*N) = Var(S)+Var(N)+Var(S)*(E(N))^2 + Var(N) * (E(S))^2.$$

This gives for the variational coefficients $v_S$ and $v_N$, i.e. the standard deviation divided by the mean

$$v_{SN}^2 = v_S*v_N+v_S^2 + v_N^2 > v_N^2.$$

So it is always favorable to analyze the number of accidents rather than the number of fatalities, since the spread of the first quantity is always smaller.

6.    **Evaluation of accident frequencies**

It is well known [6], that for a Poisson process the number of events in a time interval follows a Poisson distribution, and that, if the number of events is known (which is the case in accident evaluation) then the position of the points is independently and uniformly distributed of the interval. Also, the Poisson process can be generalized from time as a measure as exposure to other measures of exposure, e. g. train kilometer.

So if we are given two counts x and y of a particular class of event and two corresponding measures of exposure $t_1$ and $t_2$, then the probability, that a particular event falls into the first exposure interval, is given under $H_0$ by the ratio $t_1/(t_1+t_2)$. So this experiment is a Bernoulli experiment. And as the events are independent, its repetition forms a Binomial experiment. In other words, given the realizations x and y, the conditional distribution of the number of events X|x+y, that occurred in the first exposure interval, is a Binomial distribution. This is similar to Andrasik's approach, but he includes accident severities as well. So given the number of events and the operational exposure of two periods, an exact test can be derived from this result for any number of counts and non-zero operational exposures. This is known as the RateRatio test [11], which is also implemented for the statistics package R. It is known that by its construction [12] that the RateRatio test is uniformly most powerful (UMP), so for fixed error probability of the first kind, the RateRatio test has the smallest error probability of the second kind among all tests.

It should be noted that the RateRatio test is very versatile and may answer all questions asked by ERA, e. g.

- Comparison of safety performance of an operator in a test period compared to a reference period
- Comparison of safety performance of two operators in the same period
- Comparison of safety performance of two operators in different periods
- Etc.

We give a few examples to underpin this

- On EU level in 2015 there were 4 fatal train collisions and derailments, in 2016 there were 6 (similar train-km). The p value under $H_0$ is 0.38.
- An operator had no significant accident in a reference period of 4 years, and one in the following year (traffic has increased 5% p. a.). The corresponding p value is 0.22.

- Operator A has had 17 SPADs over 3 years, while operator B had 20 over 2 years (A has 1 billion train-km, B 0.7 billion train-km over the period). The p value is 0.08.

Finally, we need to choose the p-levels for our decision process. A simple choice for comparison with Andrasik are the same thresholds as BAV, this would mean p-values of 0.1 and 0.25 respectively.

At a first glance it is surprising that the differences are quite small, the deviations are highlighted in bold in the table 3.

| Number of events in reference period | Number of events in target period | | | | | | |
|---|---|---|---|---|---|---|---|
| | 0 | 1 | 2 | 3 | 4 | 5 | 6 | 7 |
| 0 | 1,000 | * **0,200** | + **0,040** | + 0,008 | | | | |
| 1 | 1,000 | 0,360 | * 0,104 | + 0,027 | + 0,007 | | | |
| 2 | 1,000 | 0,488 | * 0,181 | + 0,058 | + 0,017 | + 0,005 | | |
| 3 | 1,000 | 0,590 | **0,263** | + **0,099** | + 0,033 | + 0,010 | + 0,003 | |
| 4 | 1,000 | 0,672 | 0,345 | * 0,148 | + 0,056 | + 0,02 | + 0,006 | |
| 5 | 1,000 | 0,738 | 0,423 | * 0,203 | + 0,09 | + 0,033 | + 0,012 | + 0,004 |

Table 3  p-levels for the rate-ration test – difference with Andrasik decisions marked in bold.

Most notable are the differences when there are zero events in the 4 years reference period, but one or two events in the 1-year test period. BAV experts see no deterioration, if in the fifth year a single accident occurs, and only a potential deterioration, if two accidents occur. The RateRatio test is more restrictive.

So, we can conclude that both procedures arrive at very similar results, which differ only in particular corner cases. But with some hindsight this is clear, as they are both based on the same model, but differ only by the prior distribution. BAV has performed extensive evaluation on real as well as simulation data and has come to similar conclusions.

However, the approach of Andrasik using Bayesian methods seems to be more flexible as it takes into account also severities, which is surprising. But taking a closer look at the assumptions it becomes clear that the approach of Andrasik is here only an approximation. Formula (1) assumes that not only the events but also the fatalities are independent as the Binomial distribution takes the fatalities and severities as arguments. This is clearly not true in the general case e. g. if there is a multi-fatality accident all fatalities happen at the same time and not independently over the exposure interval. The approximation may work quite well if there are only accidents with few fatalities per event, but clearly not generally. However, it can be concluded that under the independence assumptions the p-value under this approximation is always less that the exact p-values. This means that the decision by the Andrasik approach will always be more conservative than by the exact test (which is not explicitly known). In addition, the prior distribution might influence the results, if it is not properly chosen or does not reflect the reality. This may be satisfactory for a national safety authority but not for an operator.

We may illustrate this by the example from Evans discussed already above: assume for an operator in one year 3 fatalities with level crossing accidents were reported, in the next year 11 (assuming similar operation and traffic). The RateRatio test leads to p=0.03, which would lead to the conclusion that the safety level has deteriorated. But in both years the same number of accidents has occurred – only with different severity.

This effect is particularly worse for small numbers of events. Assume that in 4 years an operator had no accident and in the next year one. RateRatio test would conclude a potential deterioration and the Andrasik approach no deterioration. But now assume that there were 2, 3, or even more fatalities in the accident. Even for two fatalities the RateRatio test applied to the accidents would lead to p=0.05 and would already conclude that there is a deterioration, while the Andrasik approach would judge it as a potential deterioration. However, with three or more fatalities both test procedures would judge a deterioration of safety level. This example also shows that the procedure would particularly discriminate against small operators.

This is caused by the following facts:

The Andrasik test is based on

- severities of accidents and the number of accidents
- combines fatalities and sever injuries as independent values, although they are not
- use a prior distribution which might influence the result for small numbers of events.

**7. Conclusions**

In this paper, we have shown that there is an exact uniformly most powerful test, the RateRatio test applied to numbers of accidents, which serves ERA's intent to compare safety levels of operators. It is applicable for any data sets with nonzero operation data and can thus be used universally. Also, it gives very similar results as a Bayesian variant, that has been calibrated on expert's judgement.

We have also shown that integrating severities into the evaluation is statistically complex and may bias the results. For one particular approach it was shown explicitly that it leads to overly conservative results.

The conclusion is that this test should also be fixed in the legal text to give legal certainty.